\documentclass[twocolumn,pra,aps,showpacs]{revtex4}

\usepackage{mathptmx}
\usepackage{subfigure}
\usepackage{psfrag,graphicx}
\usepackage{dcolumn}
\usepackage{amsmath,amssymb}
\usepackage{bm}
\usepackage{color}
\usepackage{latexsym}
\usepackage{epstopdf}
\usepackage{color}
\usepackage[english]{babel}
\usepackage{latexsym}
\usepackage{psfrag,graphicx}
\usepackage{subfigure}
\usepackage{amsmath}
\usepackage{amssymb}
\usepackage{amsfonts}
\usepackage{bm}
\usepackage{natbib}
\usepackage{epstopdf}
\usepackage{appendix}

\definecolor{mygrey}{gray}{0.35}
\definecolor{myblue}{rgb}{0.2,0.2,0.8}
\definecolor{myzard}{cmyk}{0,0,0.05,0}
\definecolor{mywhite}{rgb}{1,1,1}
\definecolor{mywhite}{rgb}{1,1,1}
\definecolor{myred}{rgb}{1,0.,0.3}

\usepackage[colorlinks=true,citecolor=myblue,linkcolor=myred]{hyperref}

\def\ba{\begin{align}}
\def\enda{\end{align}}
\def\bi{\begin{itemize}}
\def\ei{\end{itemize}}

\def\be{\begin{equation}}
\def\ee{\end{equation}}
\def\bea{\begin{eqnarray}}
\def\eea{\end{eqnarray}}
\def\bse{\begin{subequations}}
\def\ese{\end{subequations}}

\begin{document}
\title{High-precision force sensing using a single trapped ion}
\author{Peter A. Ivanov}
\affiliation{Department of Physics, St. Kliment Ohridski University of Sofia, James Bourchier 5 blvd, 1164 Sofia, Bulgaria}
\author{Nikolay V. Vitanov}
\affiliation{Department of Physics, St. Kliment Ohridski University of Sofia, James Bourchier 5 blvd, 1164 Sofia, Bulgaria}

\author{Kilian Singer}
\affiliation{Experimentalphysik I, Universit\"at Kassel, Heinrich-Plett-Str. 40, D-34132 Kassel, Germany}

\date{\today}

\begin{abstract}
We introduce quantum sensing schemes for measuring very weak forces with a single trapped ion.
They use the spin-motional coupling induced by the laser-ion interaction to transfer the relevant force information to the spin-degree of freedom.
Therefore, the force estimation is carried out simply by observing the Ramsey-type oscillations of the ion spin states.
Three quantum probes are considered, which are represented by systems obeying the Jaynes-Cummings, quantum Rabi (in 1D) and Jahn-Teller (in 2D) models. By using dynamical decoupling schemes in the Jaynes-Cummings and Jahn-Teller models, our force sensing protocols can be made robust to the spin dephasing caused by the thermal and magnetic field fluctuations. In the quantum-Rabi probe, the residual spin-phonon coupling vanishes, which makes this sensing protocol naturally robust to thermally-induced spin dephasing. We show that the proposed techniques can be used to sense the axial and transverse components of the force with a sensitivity beyond the yN $/\sqrt{\text{Hz}}$ range, i.e. in the xN$ /\sqrt{\text{Hz}}$ (xennonewton, $10^{-27}$).
The Jahn-Teller protocol, in particular, can be used to implement a two-channel vector spectrum analyzer for measuring ultra-low voltages.
\end{abstract}

\maketitle

\section{Introduction}
Over the last few years, research of mechanical systems coupled to quantum two-level systems has attracted great deal of experimental and theoretical interest \cite{Treutlein2015,Kurizki2015}.
Micro- and nano-mechanical oscillators can respond to very weak electric, magnetic and optical forces, which allows one to use them as highly sensitive force detectors \cite{Stowe1997}.
For example, the cantilever with attonewton ($10^{-18}$ N) force sensitivity can be used to test the violation of Newtonian gravity at sub-millimeter length scale \cite{Geraci2008}.
With current quantum technologies coupling between a nanomechanical oscillator and a single spin can be achieved experimentally by using strong magnetic-field gradient.
Such a coupling paves the way for sensing the magnetic force associated with the single electron spin \cite{Rugar2004}.
To this end, a recent experiment demonstrated that the coherent evolution of the electronic spin of an individual nitrogen vacancy center can be used to detect the vibration of a magnetized mechanical resonator \cite{Kolkowitz2012}.

Another promising quantum platform with application in high-precision sensing is the system of laser-cooled trapped ions, which allows excellent control over the internal and motional degrees of freedom \cite{Blatt2008}.
Force sensitivity of order of 170 yN $\rm{Hz}^{-1/2}$ ($10^{-24}$ N) was reported recently with an ensemble of ions in a Penning trap \cite{Biercuk2010}.
Force measurement down to $5$ yN has been demonstrated experimentally using the injection-locking technique with a single trapped ion \cite{Knunz2010}.
Moreover, force detection with sensitivity in the range of 1 yN $\rm{Hz}^{-1/2}$ is possible for single-ion experiments based on the measurement of the ion's displacement amplitude \cite{Maiwald2009}.

In this work, we propose ion-based sensing schemes for measuring very rapidly varying forces, which follow an earlier proposal \cite{Ivanov2015} wherein the relevant force information is mapped into the spin degrees of freedom of the single trapped ion.
In contrast to \cite{Ivanov2015}, the techniques proposed here do not require specific adiabatic evolution of the control parameters but rather they rely on using Ramsey-type oscillations of the ion's spin states, which are detected via state-dependent fluorescence measurements.
Moreover, we show that by using dynamical decoupling schemes, the sensing protocols become robust against dephasing of the spin states caused by thermal and magnetic-field fluctuations.

We consider a quantum system described by the Jaynes-Cummings (JC) model which can be used as a highly sensitive quantum probe for sensing of the axial force component.
By applying an additional strong driving field \cite{Zheng,Solano2003} the dephasing of the spin states induced by the residual spin-phonon interaction can be suppressed such that the sensing protocol does not require initial ground-state cooling of the ion's vibrational state.
We show that the axial force sensing can be implemented also by using a probe represented by the quantum Rabi (QR) model.
Because of the absence of residual spin-motional coupling in this case, the force estimation is robust to spin dephasing induced by the thermal motion fluctuations.

Furthermore, we introduce a sensing scheme capable to extract the two-dimensional map of the applied force.
Here the quantum probe is represented by the Jahn-Teller (JT) model, in which the spin states are coupled with phonons in two spatial directions.
We show that the two transverse components of the force can be measured by observing simply the coherent evolution of the spin states.
In order to protect the spin coherence during the force estimation we propose a dynamical decoupling sequence composed of phonon phase-shift operators, which average to zero the residual spin-phonon interaction.

We estimate the optimal force sensitivity in the presence of motional heating and find that with current ion trap technologies force sensitivity better than 1 yN $\rm{Hz}^{-1/2}$ can be achieved.
Thus, a single trapped ion may serve as a high-precision sensor of very weak electric fields generated by small needle electrodes with sensitivity as low as 1 $\mu{\rm V}/{\rm m}$ $\rm{Hz}^{-1/2}$.

This paper is arranged as follows.
In Sec. \ref{JCsection} we describe the sensing protocol for detection of the axial component of very weak forces using a quantum probe represented by the Jaynes-Cummings and quantum Rabi models.
It is shown that by using dynamical decoupling technique the sensing protocol using Jaynes-Cummings system is immune to thermal spin dephasing.
In Sec. \ref{JTsection} we introduce a sensing scheme, which is able to detect the two components of the external force.
Finally, in Sec. \ref{C} we summarize our findings.

\section{1D Force Sensing}\label{JCsection}

\subsection{Jaynes-Cummings quantum probe}

In our model we consider a single two-state ion with a transition frequency $\omega_{0}$, in a linear Paul trap with an axial trap frequency $\omega_{z}$.
The small axial oscillation of the ion is described by the vibrational Hamiltonian $\hat{H}_{\rm ax}=\hbar\omega_{z}\hat{a}^{\dag}\hat{a}$, where $\hat{a}^{\dag}$ ($\hat{a}$) creates (annihilates) a phonon excitation.
We assume that the ion interacts with a laser field with a frequency $\omega_{\rm L}=\omega_{0}-\omega_{z}+\delta$, tuned near the red-sideband resonance with a detuning $\delta$.
The interaction Hamiltonian in the Lamb-Dicke limit and the rotating-wave approximation reads \cite{Wineland1998,Haffner2008,Schneider2012}
\begin{equation}
\hat{H}_{\rm JC}=\hbar\omega \hat{a}^{\dag}\hat{a}+\hbar\Delta\sigma_{z}+\hbar g(\sigma^{-}\hat{a}^{\dag}+\sigma^{+}\hat{a}),\label{HJC}
\end{equation}
with $\delta=\Delta-\omega$, where $\Delta$ is the effective spin frequency and $\omega$ is the effective phonon frequency.
Here, $\sigma_{x,y,z}$ are the Pauli matrices, $\sigma^{\pm}$ are the respective raising and lowering operators for the effective spin system, and $g$ determines the strength of the spin-phonon coupling.

The external time-varying force with a frequency $\omega_{d}=\omega_{z}-\omega$, e.g., $F(t)=F\cos(\omega_{d} t)$, displaces the motional amplitude of the ion oscillator along the axial direction, as described by the term
\begin{equation}
\hat{H}_{F}=\frac{z_{\rm ax}F}{2}(\hat{a}^{\dag}+\hat{a}).\label{Hf}
\end{equation}
Here $z_{\rm ax}=\sqrt{\hbar/2m\omega_{z}}$ is the spread of the zero-point wavefunction along the axial direction and $F$ is the parameter we wish to estimate.
The origin of the oscillating force can be a very weak electric field, an optical dipole force, spin-dependent forces created in a magnetic-field gradient or a Stark-shift gradient, etc.
With the term \eqref{Hf} the total Hamiltonian becomes
\begin{equation}
\hat{H}_{\rm T}=\hat{H}_{\rm JC}+\hat{H}_{F}.\label{HTJC}
\end{equation}

In the following, we consider the weak-coupling regime $g\ll \omega$, in which the phonon degree of freedom can be eliminated from the dynamics.
This can be carried out by applying the canonical transformation $\hat{U}=e^{\hat{S}}$ to $\hat{H}_{\rm T}$ (\ref{HTJC}) such that $\hat{H}_{\rm eff}^{\rm JC}=e^{-\hat{S}}\hat{H}_{\rm T}e^{\hat{S}}$ with $\hat{S}=(g/\omega)(\sigma^{+}\hat{a}-\sigma^{-}\hat{a}^{\dag})+(\Omega_{F}/g)(\hat{a}-\hat{a}^{\dag})$.
Keeping only the terms of order of $g/\omega$ we arrive at the following effective Hamiltonian (see the Appendix),
\bse
\begin{align}\label{HeffJC}
\hat{H}_{\rm eff}^{\rm{JC}} &= \hbar\widetilde{\Delta}\sigma_{z}-\hbar\Omega_{F}\sigma_{x}-\hat{H}_{\rm JC}^{\prime}, \\
\hat{H}^{\prime}_{\rm JC} &= \frac{\hbar g^{2}}{\omega}\sigma_{z}\hat{a}^{\dag}\hat{a}.
\end{align}
\ese
This result indicates that the spin-motional interaction in Eq.~\eqref{HTJC} shifts the effective spin frequency by the amount $\widetilde{\Delta}=\Delta-g^{2}/2\omega$, while the effect of the force term is to induce transitions between the spin states.
The strength of the transition is quantified by the Rabi frequency $\Omega_{F}=g z_{\rm ax}F/2\hbar\omega$, which is proportional to the applied force $F$.
Hence the force estimation can be carried out by observing the coherent evolution of the spin population that can be read out via state-dependent fluorescence.

\begin{figure}
\includegraphics[width=0.45\textwidth]{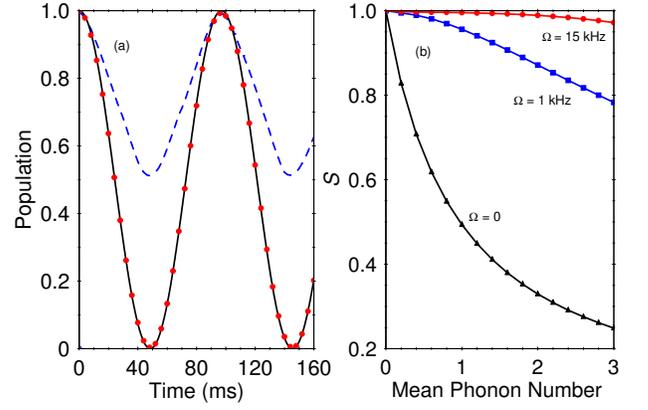}
\caption{(Color online)
a) Time-evolution of the probability to find the system in spin state $\left|\uparrow\right\rangle$ for the JC system.
We compare the probabilities derived from the original Hamiltonian (\ref{HTJC}) (dots) and the effective Hamiltonian (\ref{HeffJC}) (solid lines). We assume an initial thermal distribution with a mean phonon number $\bar{n}=1.2$.
The parameters are set to $g=4$ kHz, $\omega=170$ kHz, $\Delta=g^{2}/2\omega$, $z_{\rm ax}=14.5$ nm, $F=20$ yN and $\Omega=10$ kHz.
For the same initial state but in the absence of driving field ($\Omega=0$), the signal loses contrast (blue dashed line).
b) Contrast of the Rabi oscillations defined as $S=P_{\uparrow}(t_{2})-P_{\uparrow}(t_{1})$ with $t_{1}=\pi/2\Omega_{F}$ and $t_{2}=\pi/\Omega_{F}$ with $\Omega_{F}=60$ kHz as a function of the mean phonon number $\bar{n}$. }
\label{fig1}
\end{figure}

The last term  $\hat{H}_{\rm JC}^{\prime}$ in Eq.~\eqref{HeffJC} is the residual spin-motional coupling.
This term affects the force estimation because it can be a source of pure spin dephasing \cite{Viola1999}.
Indeed, the $\sigma_{z}$ factor in $\hat{H}_{\rm JC}^{\prime}$ induces transitions between the eigenstates $\left|\pm\right\rangle$ of the operator $\sigma_{x}$ depending on the vibrational state of the oscillator.
As long as the oscillator is prepared initially  in an incoherent vibrational state at a finite temperature this would lead to a random component in the spin energy.
As we will see below, by using  dynamical decoupling the effect of the pure spin dephasing can be reduced.

The sensing protocol starts by preparing the system in state $\hat{\rho}(0)=\left|\uparrow\right\rangle\left\langle\uparrow\right|\otimes\hat{\rho}_{\rm osc}$, where $\hat{\rho}_{\rm osc}$ stands for the initial density operator of the oscillator.
According to Eq.~\eqref{HeffJC}, the evolution of the system is driven by the unitary propagator $\hat{U}_{\rm JC}(t,0)=e^{-{\rm i}\hat{H}_{\rm eff}^{\rm JC} t/\hbar}$.
Assuming for the moment that $\hat{\rho}_{\rm osc}=|0\rangle\langle 0|$ where $|n\rangle$ is the harmonic oscillator Fock state with $n$ phonon excitations, the probability to find the system in state $\left|\uparrow\right\rangle$ is $P_{\uparrow}(t)=\cos^{2}(\Omega_{F} t)$, where for simplicity we set $\Delta=g^{2}/2\omega$, hence $\widetilde\Delta = 0$.
In this case, the effect of $\hat{H}_{\rm JC}^{\prime}$ automatically vanishes such that the signal exhibits a cosine behavior according to the effective Hamiltonian (\ref{HeffJC}).
An initial thermal phonon distribution, however, would introduce dephasing on the spin oscillations caused by thermal fluctuations.
The spin coherence can be protected, for example, by applying a sequence of fast pulses, which flip the spin states and average the residual spin-motional interaction to zero during the force estimation \cite{Yang2010}.
On the other hand, because the relevant force information is encoded in the $\sigma_{x}$ term in Eq.~\eqref{HeffJC}, continuously applying an additional strong driving field $\hat{H}_{\rm d}=\hbar\Omega\sigma_{x}$ in the same basis  \cite{Zheng,Solano2003}, such that $\hat{H}_{\rm T}\rightarrow \hat{H}_{\rm T}+\hat{H}_{\rm d}$, would not affects the force estimation but rather will suppress the effect of the residual spin-motional coupling.
Indeed, going in the interaction frame with respect to $\hat{H}_{\rm d}$, the residual spin-motional coupling becomes
\be
\hat{H}_{\rm JC}^{\prime}(t)=\frac{\hbar g^{2}}{\omega}(e^{2{\rm i}\Omega t}|+\rangle\langle -|+e^{-2{\rm i}\Omega t}|-\rangle\langle+|)\hat{a}^{\dag}\hat{a}.
\ee
The latter result indicates that the off-resonance transitions between states $\left|\pm\right\rangle$ induced by $\hat{H}_{\rm JC}^{\prime}$ are suppressed if $g^{2}/2\omega\ll\Omega$.
By separating the pulse sequences from $t=0$ to $t/2$ with a Hamiltonian $\hat{H}_{\rm T}+\hat{H}_{\rm d}$, and then from $t/2$ to $t$ with a Hamiltonian $\hat{H}_{\rm T}-\hat{H}_{\rm d}$, the spin states are protected from the thermal dephasing and the signal depends only on the Rabi frequency $\Omega_{F}$ at the final time $t$.
Note that the effect of the magnetic field fluctuations of the spin states is described by an additional $\sigma_{z}$ term in Eq.~\eqref{HeffJC}, therefore the strong driving field used here suppresses the spin dephasing caused by the magnetic-field fluctuations, as was experimentally demonstrated \cite{Timoney2011,Webster2013}.

In Fig. \ref{fig1}(a) we show the time evolution of the probability $P_{\uparrow}(t)$ for an initial thermal vibrational state.
Applying the driving field during the force estimation leads to reduction of the spin dephasing and hence protecting the contrast of the Rabi oscillations, see Fig. \ref{fig1}(b).
We note that a similar technique using a strong driving carrier field for dynamical decoupling was proposed for the implementation of a high-fidelity phase gate with two trapped ions \cite{Tan2013,Bermudez2012}.


\begin{figure}
\includegraphics[width=0.45\textwidth]{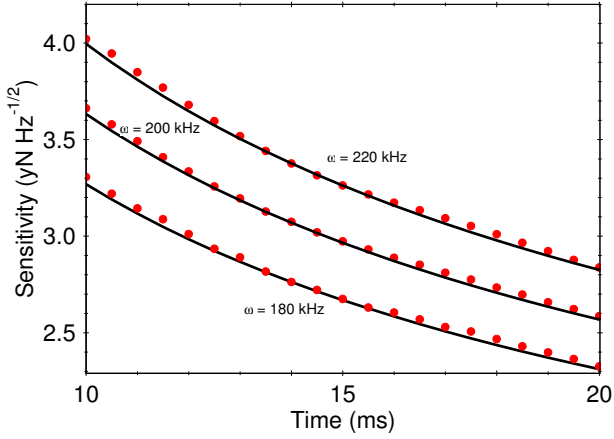}
\caption{(Color online) The sensitivity of the force measurement versus time $t$ for various values of $\omega$.
We assume an initial thermal vibrational state with a mean phonon number $\bar{n}=1$.
The solid lines represent the analytical result given by Eq. (\ref{Force_Sens}) while the dots are the exact numerical solution with the Hamiltonian (\ref{HTJC}) including the strong driving term.
The other parameters are set to $g=4$ kHz and $\Omega=7$ kHz. }
\label{fig2}
\end{figure}

The shot-noise-limited sensitivity for measuring $\Omega_{F}$ is
\begin{equation}
\delta\Omega_{F}=\frac{\Delta P_{\uparrow}(t)}{\frac{\partial P_{\uparrow}(t)}{\partial\Omega_{F}}\sqrt{\nu}},\label{SN}
\end{equation}
where $\Delta P_{\uparrow}(t)$ stands for the variance of the signal and $\nu=T/\tau$ is the repetition number. Here $T$ is the total experimental time, and the time $\tau$ includes the evolution time as well as the preparation and measurement times.
Because our technique relies on state-projective detection, such that the preparation and measurement times are much smaller than the other time scale, we assume $\tau\approx t$.
From Eq. (\ref{SN}) we find that the sensitivity, which characterizes the minimal force difference that can be discriminated within a total experimental time of 1 s, is
\begin{equation}
F_{\rm min}\sqrt{T}=\frac{\hbar\omega}{gz_{\rm ax}\sqrt{t}}\label{Force_Sens}.
\end{equation}

In Fig. \ref{fig2} we show the sensitivity of the force estimation versus time $t$ for different frequencies $\omega$ assuming an initial thermal  vibrational state.
For an evolution time of $20$ ms, force sensitivity of $2$ yN $\rm{Hz}^{-1/2}$ can be achieved.

Let us now estimate the effect of the motional heating which limits the force estimation.
Indeed, the heating of the ion motion causes damping of the signal, which leads to \cite{Wineland1998}
\be
P_{\uparrow}(t)=\frac{1}{2} [1+e^{-\gamma t}\cos(2\Omega_{F} t)],
\ee
 where $\gamma$ is the decoherence rate.
We assume that $\gamma\sim\langle \dot{n}_{\rm ax}\rangle$ where $\langle \dot{n}_{\rm ax}\rangle$ stands for the axial ion's heating rate.
Thus, the optimal force sensitivity is \cite{Huelga1997}
\begin{equation}
F_{\rm min}\sqrt{T}=\frac{\hbar\omega}{gz_{\rm ax}}\sqrt{2\langle \dot{n}_{\rm ax}\rangle e}.\label{Fopt}
\end{equation}
Using the parameters in Fig. \ref{fig2} with $\omega=180$ kHz and assuming $\langle \dot{n}_{\rm ax}\rangle=0.01$ $\rm{ms}^{-1}$ we estimate force sensitivity of 2.4 yN $\rm{Hz}^{-1/2}$.
For a cryogenic ion trap with heating rate in the range of $\langle \dot{n}_{\rm ax}\rangle=1$ ${\rm s}^{-1}$ and evolution time of $t=500$ ms, the force sensitivity would be 0.8 yN $\rm{Hz}^{-1/2}$.

\begin{figure}

\includegraphics[width=0.45\textwidth]{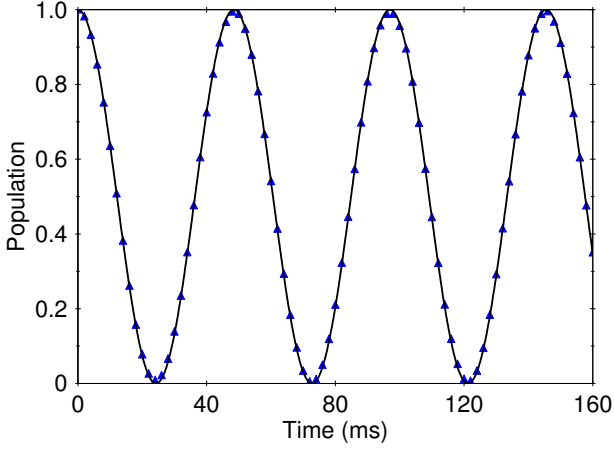}
\caption{(Color online) Time-evolution of the probability to find the system in spin state $\left|\uparrow\right\rangle$ for the QR system.
We assume an initial thermal vibrational state with a mean phonon number $\bar{n}=1.2$. Due to the absence of residual spin-motion coupling the Rabi oscillations are robust with respect to the spin dephasing caused by the thermal fluctuations. We compare the probability derived from the Hamiltonian $\hat{H}_{\rm T}=\hat{H}_{\rm QR}+\hat{H}_{F}$ with the analytical solution $P_{\uparrow}=\cos^{2}(2\Omega_{F})$. The parameters are set to $g=4$ kHz, $\omega=170$ kHz, $z_{\rm ax}=14.5$ nm, $F=20$ yN.}
\label{fig3}
\end{figure}

\subsection{Quantum Rabi model}

An alternative approach to sense the axial component of the force is to use a probe described by the quantum Rabi model,
\begin{equation}
\hat{H}_{\rm QR}=\hbar\omega \hat{a}^{\dag}\hat{a}+\hbar g\sigma_{x}(\hat{a}^{\dag}+\hat{a}),
\end{equation}
which includes it the counter-rotating wave terms.
This Hamiltonian can be implemented by using a bichromatic laser field along the axial direction \cite{Pedernales2015}.
In the weak-coupling regime, $g\ll\omega$, we find by using the unitary transformation $\hat{U}=e^{\hat{S}}$ with $\hat{S}=-(g/\omega)\sigma_{x}(\hat{a}^{\dag}-\hat{a})-(2\Omega_{F}/g) (\hat{a}^{\dag}-\hat{a})$ that (see the Appendix)
\begin{equation}
\hat{H}_{\rm eff}^{\rm QR}=-2\hbar\Omega_{F}\sigma_{x}.\label{HQREff}
\end{equation}
In contrast to Eq.~\eqref{HeffJC}, now the effective Hamiltonian (\ref{HQREff}) does not contain an additional residual spin-motional coupling, which implies that the spins are immune to dephasing caused by the thermal motion fluctuations, see Fig. \ref{fig3}.
Thereby the force estimation can be carried out without additional strong driving field. We find that the optimal force sensitivity is similar to Eq. (\ref{Fopt}) but with extra factor of 2 in the denominator,
\begin{equation}
F_{\rm min}\sqrt{T}=\frac{\hbar\omega}{2gz_{\rm ax}}\sqrt{2\langle \dot{n}_{\rm ax}\rangle e}.\label{Fopt-qe}
\end{equation}

Up to now we have considered probes that are responsive only to the axial component of the force.
In the following we propose a sensing technique that can be used to detect the two transverse components of the time-varying external force.

\section{Jahn-Teller quantum probe}\label{JTsection}

In conventional ion trap sensing methods, the information on the force direction can be extracted by using the three spatial vibrational modes of the ion \cite{Maiwald2009,Munro2003}.
Such an experiment requires an independent measurement of the displacement amplitudes in each vibrational mode, which, however, increases the complexity of the measurement procedure and can lead to longer total experimental times.
Here we show that by utilizing the laser-induced coupling between the spin states and the transverse ion oscillation we are able to detect the transverse components of the force by observing simply the coherent evolution of the spin states.

Indeed, let us consider the case in which the small transverse oscillations of the ion with a frequency $\omega_{\rm t}$ described by the Hamiltonian $\hat{H}_{\rm t}=\hbar\omega_{\rm t}(\hat{a}_{x}^{\dag}\hat{a}_{x}+\hat{a}_{y}^{\dag}\hat{a}_{y})$ are coupled with the spin states via Jahn-Teller interaction.
Such a coupling can be achieved by using bihromatic laser fields with frequencies $\omega_{b,r}=\omega_{0}\pm(\omega_{\rm t}-\omega)$ tuned respectively near the blue- and red-sideband resonances, with a detuning $\omega$, which excite the transverse $x$ and $y$ vibrational modes of the trapped ion.
The interaction Hamiltonian of the system is given by \cite{Porras2012,Ivanov2013}
\begin{equation}
\hat{H}_{\rm JT}=\hbar\omega(\hat{a}^{\dag}_{x}\hat{a}_{x}+\hat{a}^{\dag}_{y}\hat{a}_{y})+\hbar g\sigma_{x}(\hat{a}^{\dag}_{x}+\hat{a}_{x})+\hbar g\sigma_{y}(\hat{a}^{\dag}_{y}+\hat{a}_{y}).\label{HJT}
\end{equation}
Here $\hat{a}^{\dag}_{\beta}$ and $\hat{a}_{\beta}$ are the creation and annihilation operators of phonon excitations along the transverse direction ($\beta=x,y$) with an effective frequency $\omega$.
The last two terms in Eq.~\eqref{HJT} describe the Jahn-Teller $E\otimes e$ spin-phonon interaction with a coupling strengths $g$.
In the following, we assume that a classical oscillating force with a frequency $\omega_{d}=\omega_{\rm t}-\omega$  displaces the vibrational amplitudes along the transverse $x$ and $y$ directions of the quantum oscillator described by
\begin{equation}
\hat{H}_{\vec{F}}=\frac{z_{\rm t}F_{x}}{2}(\hat{a}^{\dag}_{x}+\hat{a}_{x})+\frac{z_{\rm t}F_{y}}{2}(\hat{a}^{\dag}_{y}+\hat{a}_{y}),\label{Hpert}
\end{equation}
where $z_{\rm t}=\sqrt{\hbar/2m\omega_{\rm t}}$ is the size of the transverse ion's harmonic oscillator ground-state wavefunction.
$F_{x}$ and $F_{y}$ are the two transverse components of the force we wish to estimate.
With the perturbation term \eqref{Hpert} the total Hamiltonian becomes
\begin{equation}
\hat{H}_{\rm T}=\hat{H}_{\rm JT}+\hat{H}_{\vec{F}}.\label{HT}
\end{equation}
Assuming the weak-coupling regime, $g\ll\omega$, the two phonon modes are only virtually excited.
After performing the canonical transformation $\hat{U}=e^{\hat{S}}$ of $\hat{H}_{\rm T}$ \eqref{HT}, where
\begin{equation}
\hat{S}=(\hat{a}_{x}-\hat{a}_{x}^{\dag})\left(\frac{g}{\omega}\sigma_{x}+\frac{\Omega_{x}}{g}\right)+
(\hat{a}_{y}-\hat{a}_{y}^{\dag})\left(\frac{g}{\omega}\sigma_{y}+\frac{\Omega_{y}}{g}\right),
\end{equation}
we obtain the following effective Hamiltonian (see the Appendix)
\begin{equation}
\hat{H}_{\rm eff}^{\rm{JT}}=-\hbar\Omega_{x}\sigma_{x}-\hbar\Omega_{y}\sigma_{y}+\hat{H}^{\prime}_{\rm JT}.\label{Heff}
\end{equation}
Here $\Omega_{x,y}=gz_{\rm t}F_{x,y}/\hbar\omega$ are the respective driving Rabi frequencies of the transition between spin states $\left|\uparrow\right\rangle$ and $\left|\downarrow\right\rangle$.
The last term in Eq.~\eqref{Heff} is the residual spin-phonon interaction described by
\begin{equation}
\hat{H}^{\prime}_{\rm JT}=2{\rm i}\frac{\hbar g^{2}}{\omega}\sigma_{z}(\hat{a}_{x}^{\dag}\hat{a}_{y} -\hat{a}_{x}\hat{a}_{y}^{\dag}),\label{JT_res}
\end{equation}
which can be a source of thermal spin dephasing as long as the two phonon modes are prepared in initial thermal vibrational states.

\begin{figure}
\includegraphics[width=0.45\textwidth]{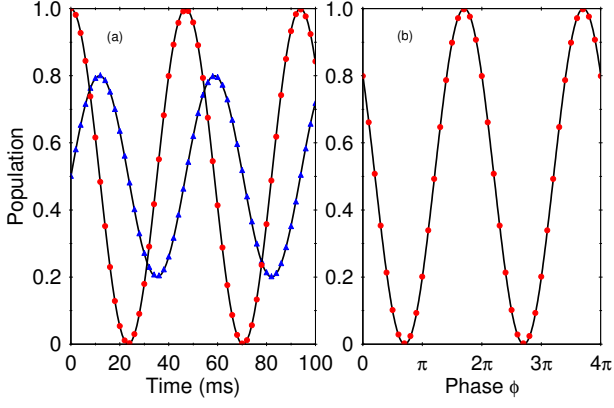}
\caption{(Color online)
a) Time-evolution of the probability to find the system in spin state $\left|\uparrow\right\rangle$ for the JT system.
We compare the probability calculated from the Hamiltonian (\ref{HT}) assuming the initial states $\left|\psi(0)\right\rangle=\left|\uparrow\right\rangle\left|0_{x},0_{y}\right\rangle$ (red dots) and $\left|\psi(0)\right\rangle=2^{-1/2}(\left|\uparrow\right\rangle+\left|\downarrow\right\rangle)|0_{x},0_{y}\rangle$ (blue triangles) with those given by the effective Hamiltonian \eqref{Heff} (solid lines).
The parameters are set to $g=4$ kHz, $\omega=170$ kHz, $z_{\rm t}=12$ nm, $F_{x}=20$ yN and $F_{y}=15$ yN.
b) Oscillations of the signal for fixed $t$ as a function of the phase $\phi$ for an initial superposition spin state.}
\label{fig4}
\end{figure}

The two-dimensional force sensing protocol starts by preparing the system in state $|\psi(0)\rangle=(c_{\uparrow}(0)\left|\uparrow\right\rangle+c_{\downarrow}(0)\left|\downarrow\right\rangle)\otimes\left|0_{x},0_{y}\right\rangle$, where $c_{\uparrow,\downarrow}(0)$ are the respective initial spin probability amplitudes and $\left|n_{x},n_{y}\right\rangle$ stands for the Fock state with $n_{\beta}$ excitations in each phonon mode.
According to the effective Hamiltonian (\ref{Heff}) the evolution of the system is driven by the free propagator $\hat{U}_{\rm JT}=e^{-{\rm i}\hat{H}_{\rm eff}^{\rm JT}t/\hbar}$.
Neglecting the residual spin-motional coupling (\ref{JT_res}) the propagator reads
\begin{equation}
\hat{U}_{\rm JT}^{0}(t,0) =\left[\begin{array}{cc}
a & b \\ -b^{*} &  a^{*}
\end{array}\right].\label{U}
\end{equation}
Here $a=\cos(\widetilde{\Omega} t)$ and $b={\rm i}e^{-\rm{i}\xi}\sin(\widetilde{\Omega} t)$ are the Cayley-Klein parameters, which depend on the rms Rabi frequency $\widetilde{\Omega}=\frac{g z_{\rm t}}{\hbar\omega}|\vec{F}_{\perp}|$, which is proportional to the magnitude of the force $|\vec{F}_{\perp}|=\sqrt{F_{x}^{2}+F_{y}^{2}}$.
In addition to $|\vec{F}_{\perp}|$, we introduce the relative amplitude parameter $\xi=\tan^{-1}\left(\frac{F_{y}}{F_{x}}\right)$.
Assuming an initial state with $c_{\uparrow}(0)=1$, $c_{\downarrow}(0)=0$, the respective probability to find the system in state $\left|\uparrow\right\rangle$ is $P_{\uparrow}(t)=\cos^{2}(\widetilde{\Omega} t)$,
 which implies that the Rabi oscillations depends only on the magnitude of the force, see Fig. \ref{fig4}(a).
Using Eq. (\ref{SN}) we find that the shot-noise-limited sensitivity for measuring the magnitude of the force is given by
\begin{equation}
|\vec{F}_{\perp}|_{\rm min}\sqrt{T}=\frac{\hbar\omega}{2gz_{\rm t}\sqrt{t}}.
\end{equation}
In the presence of motional heating of both vibrational modes, the signal is damped with decoherence rate $\gamma\sim\langle \dot{n}_{x}\rangle+\langle \dot{n}_{y}\rangle$, where $\langle \dot{n}_{\beta}\rangle$ is the heating rate along the $\beta$ spatial direction.
Therefore we find that the optimal force sensitivity is
\begin{equation}
|\vec{F}_{\rm\perp}|_{\rm min}\sqrt{T}=\frac{\hbar\omega}{2gz_{\rm t}}\sqrt{2(\langle \dot{n}_{x}\rangle+\langle \dot{n}_{y}\rangle) e}.
\end{equation}
It is important that due to the strong transverse confinement the sensing scheme for measuring $|\vec{F}_{\perp}|$ is less sensitive to the ion's heating \cite{Turchette2000,Zhu2006}. Using the parameters in Fig. \ref{fig4} and assuming $\langle \dot{n}_{x}\rangle=\langle \dot{n}_{y}\rangle=1$ $\rm{s}^{-1}$ we estimate force sensitivity of 0.6 yN $\rm{Hz}^{-1/2}$.

In order to detect the parameter $\xi$ we prepare the spin state in an initial superposition state with $c_{\uparrow}(0)=1/\sqrt{2}$ and $c_{\downarrow}(0)=e^{{\rm i}\phi}/\sqrt{2}$. Then the probability oscillates with time as
\begin{eqnarray}\label{2d-Ramsey}
P_{\uparrow}(t) = \frac{1}{2}\left[ 1+\sin(\xi-\phi)\sin(2\widetilde{\Omega} t) \right].
\end{eqnarray}
Hence, for fixed evolution time $t$, the Ramsey oscillations versus the phase $\phi$ provide a measure of the relative phase $\xi$, see Fig. \ref{fig4}(b).

In fact, Eq.~\eqref{2d-Ramsey} allows one to determine both the magnitude of the force $|\widetilde{\Omega}|$ and the mixing parameter $\xi$ from the same signal when plotted vs the evolution time $t$:  $|\widetilde{\Omega}|$ is related to the oscillation frequency and $\xi$ to the oscillation amplitude.
The parameter $\xi$ can be determined also by varying the externally controlled superposition phase $\phi$, until the oscillation amplitude vanishes at some value $\phi_0$; this signals the value $\xi=\phi_0$ (modulo $\pi$).

Finally, we discuss the dynamical decoupling schemes, which can be used to suppress the effects of the term $\hat{H}_{\rm JT}^{\prime}$ (\ref{JT_res}) during the force estimation.
In that case, applying continuous driving field, e.g., along the $\sigma_{x}$ direction, would reduce the thermal fluctuation induced by $\hat{H}_{\rm JT}^{\prime}$, but additionally, the relevant force information, which is encoded in the $\sigma_{y}$ term in \eqref{Heff}, will be spoiled.
Here we propose an alternative dynamical decoupling scheme, which follows the Carr-Purcell-Meiboom-Gill (CPMG) pulse sequence \cite{Carr1954,Meiboom1958}, in which, however, the single instantaneous $\pi$ pulse is replaced by the phonon phase-flip operator $\hat{R}_{\pi}=e^{{\rm i}\pi\hat{a}_{x}^{\dag}\hat{a}_{x}}$.
Such a phonon phase shift $\Delta\omega_{x}\tau=\pi$ can be achieved by switching the RF potential of the trap by the fixed amount $\Delta\omega_{x}$ for a time $\tau$ \cite{Singer2010}.
The effect of $\hat{R}_{\pi}$ is to change the sign of the $\hat{H}_{\rm JT}^{\prime}$ such that $\hat{R}^{\dag}_{\pi}\hat{H}_{\rm JT}^{\prime}\hat{R}_{\pi}=-\hat{H}_{\rm JT}^{\prime}$ but it leaves the other part of the Hamiltonian (\ref{Heff}) unaffected.
Using that the pulse sequence $\hat{U}_{1}=\hat{R}_{\pi}\hat{U}_{\rm JT}\hat{R}_{\pi}\hat{U}_{\rm JT}$ eliminates the residual spin-phonon coupling in the first order of the interaction time $t$,
a high-order reduction can be achieved by the recursion $\hat{U}_{n}=\hat{R}_{\pi}\hat{U}_{n-1}\hat{R}_{\pi}\hat{U}_{n-1}$, which eliminates the spin-phonon coupling up to $n$th order in $t$.

\section{Summary and Outlook}\label{C}

We have proposed quantum sensing protocols, which rely on mapping the relevant force information onto the spin degrees of freedom of the single trapped ion.
The force sensing is carried out by observing the Ramsey-type oscillations of the spin states, which can be detected via state-dependent fluorescence.
We have considered quantum probes represented by the JC and QR systems, which can be used to sense the axial component of the force.
We have shown that when using a JC system as a quantum probe, one can apply dynamical decoupling schemes to suppress the effect of the spin dephasing during the force estimation.
When using a QR system as a probe, the absence of a residual spin-phonon coupling makes the sensing protocol robust to thermally-induced spin dephasing.
Furthermore, we have shown that the transverse-force direction can be measured by using a system described by the JT model, in which the spin states are coupled with the two spatial phonon modes. Here the information of the magnitude of the force and the relative ratio can be extracted by observing the time evolution of the respective ion's spin states, which simplify significantly the experimental procedure.

Tuning the trap frequencies over the broad range, the force sensing methods proposed here can be employed to implement a spectrum analyzer for ultra-low voltages.
Moreover, because in the force-field direction sensing the mutual ratio can be additionally estimated our method can be used to implement a two-channel vector spectrum analyzer.
Finally, the realization of the proposed force sensing protocols are not restricted only to trapped ions but could be implemented with other quantum optical setups such as cavity-QED \cite{Dimer2007} or circuit-QED systems \cite{Ballester2015}.


\appendix

\section{Elimination of the vibrational degree of freedom\label{Dsignal}}

Let us make the canonical transformation of Hamiltonian $\hat{H}=\hat{H}_{0}+\hat{H}_{\rm int}$,
\begin{eqnarray}
\hat{H}_{\rm eff}&=&e^{-\hat{S}}\hat{H}e^{\hat{S}}=\hat{H}_{0}+\hat{H}_{\rm int}+[\hat{H}_{0},\hat{S}]+[\hat{H}_{\rm int},\hat{S}]\notag\\
&&+\tfrac{1}{2}[[\hat{H}_{0},\hat{S}],\hat{S}]+\tfrac{1}{2}[[\hat{H}_{\rm int},\hat{S}],\hat{S}]+\ldots.\label{H_CT}
\end{eqnarray}
Our goal is to choose $\hat{S}$ in a such a way that all terms of order $g$ in $\hat{H}_{\rm eff}$ are canceled and the first term describing the spin-boson interaction is of order $g^{2}/\omega$. If we determine $\hat{S}$ by the condition
\begin{equation}
\hat{H}_{\rm int}+[\hat{H}_{0},\hat{S}]=0,\label{condS}
\end{equation}
then the effective Hamiltonian becomes
\begin{equation}
\hat{H}_{\rm eff}\approx \hat{H}_{0}+\tfrac{1}{2}[\hat{H}_{\rm int},\hat{S}].
\end{equation}
Let us consider the time-dependent operator $\hat{S}(t)=e^{{\rm i}\hat{H}_{0}t/\hbar}\hat{S}e^{-{\rm i}\hat{H}_{0}t/\hbar}$, which obeys the Heisenberg equation ${\rm i}\hbar \dot{\hat{S}}(t)=[\hat{S}(t),\hat{H}_{0}]$. Using Eq. (\ref{condS}) we arrive at the equation
\begin{equation}
{\rm i}\hbar\dot{\hat{S}}(t)=\hat{H}_{\rm int}(t),\label{S}
\end{equation}
where $\hat{H}_{\rm int}(t)=e^{{\rm i} \hat{H}_{0}t/\hbar}\hat{H}_{\rm int}e^{-{\rm i}\hat{H}_{0}t/\hbar}$. Solving Eq. (\ref{S}) we determine the desired operator $\hat{S}$.

\subsection{Jaynes-Cummings model}

We identify $\hat{H}_{0}=\hbar\omega \hat{a}^{\dag}\hat{a}$ and $\hat{H}_{\rm int}=\hbar g(\sigma^{-}\hat{a}^{\dag}+\sigma^{+}\hat{a})+\frac{z_{\rm ax}F}{2}(\hat{a}^{\dag}+\hat{a})$. Using Eq. (\ref{S}) we obtain
\begin{equation}
\hat{S}=\frac{g}{\omega}(\sigma^{+}\hat{a}-\sigma^{-}\hat{a}^{\dag})+\frac{z_{\rm ax}F}{2\hbar\omega}(\hat{a}-\hat{a}^{\dag}),
\end{equation}
which fulfills the condition (\ref{condS}). For the effective Hamiltonian we derive
\begin{eqnarray}
\hat{H}_{\rm eff}&=&\hbar\omega \hat{a}^{\dag}\hat{a}+\hbar\left(\Delta-\frac{g^{2}}{2\omega}\right)\sigma_{z}-\hbar\Omega_{F}\sigma_{x}\notag\\
&&-\frac{\hbar g^{2}}{\omega}\sigma_{z}\hat{a}^{\dag}\hat{a}-\frac{\hbar g^{2}}{2\omega}-\frac{z_{\rm ax}^{2}F^{2}}{4\hbar\omega}+\hat{H}^{\prime},
\end{eqnarray}
where $\Omega_{F}=gz_{\rm ax}F/2\hbar\omega$ is the Rabi frequency and $\hat{H}^{\prime}=\frac{1}{3}[[\hat{H}_{\rm int},\hat{S}],\hat{S}]+\ldots$ contains the higher-order terms in (\ref{H_CT}). We find
\begin{eqnarray}
\frac{1}{3}[[\hat{H}_{\rm int},\hat{S}],\hat{S}]&=&\frac{2g^{2}z_{\rm ax}F}{3\omega^{2}}\sigma_{z}(\hat{a}^{\dag}+\hat{a})-\frac{4\hbar g^{3}}{3\omega^{2}}(\sigma^{-}\hat{a}^{\dag}+\sigma^{+}\hat{a})\notag\\
&&-\frac{4\hbar g^{3}}{3\omega^{2}}(\sigma^{-}\hat{a}^{\dag}\hat{a}^{\dag}\hat{a}+\sigma^{+}\hat{a}^{\dag}\hat{a}\hat{a}).
\end{eqnarray}
As long as $g/\omega \ll 1$ the higher-order terms can be neglected and thus the lowest-order effective Hamiltonian is given by Eq. (\ref{HeffJC}).

\subsection{Quantum Rabi Model}
Here the interaction Hamiltonian is $\hat{H}_{\rm int}=\hbar g\sigma_{x}(\hat{a}^{\dag}+\hat{a})+\frac{z_{\rm ax}F}{2}(\hat{a}^{\dag}+\hat{a})$ and the canonical transformation is given by the operator
\begin{equation}
\hat{S}=\frac{g}{\omega}\sigma_{x}(\hat{a}-\hat{a}^{\dag})+\frac{z_{\rm ax}F}{2\hbar\omega}(\hat{a}-\hat{a}^{\dag}).
\end{equation}
The effective Hamiltonian is
\begin{equation}
\hat{H}_{\rm eff}=\hbar\omega \hat{a}^{\dag}\hat{a}-2\hbar\Omega_{F}\sigma_{x}-\frac{\hbar g^{2}}{\omega}-\frac{(z_{\rm ax}F)^{2}}{4\hbar\omega}.
\end{equation}
Remarkably, due to the equality $[[\hat{H}_{\rm int},\hat{S}],\hat{S}]=0$ all higher-order terms in Eq. (\ref{H_CT}) vanish.

\subsection{Jahn-Teller Model}

Following the same procedure we have
\begin{eqnarray}
&&\hat{H}_{0}=\hbar\omega(\hat{a}_{x}^{\dag}\hat{a}_{x}+\hat{a}_{y}^{\dag}\hat{a}_{y}),\notag\\
&&\hat{H}_{\rm int}=\hbar g\sigma_{x}(\hat{a}_{x}^{\dag}+\hat{a}_{x})+\hbar g\sigma_{y}(\hat{a}_{y}+\hat{a}_{y})+\frac{z_{\rm t}F_{x}}{2}(\hat{a}_{x}^{\dag}+\hat{a}_{x})\notag\\
&&+\frac{z_{\rm t}F_{y}}{2}(\hat{a}^{\dag}_{y}+\hat{a}_{y}).
\end{eqnarray}
In this case the canonical transformation is represented by the operator
\begin{eqnarray}
\hat{S}&=&\frac{g}{\omega}\sigma_{x}(\hat{a}_{x}-\hat{a}_{x}^{\dag})+\frac{g}{\omega}\sigma_{y}(\hat{a}_{y}-\hat{a}_{y}^{\dag})
+\frac{z_{\rm t}F_{x}}{2\hbar\omega}(\hat{a}_{x}-\hat{a}_{x}^{\dag})\notag\\
&&+\frac{z_{\rm t}F_{y}}{2\hbar\omega}(\hat{a}_{y}-\hat{a}_{y}^{\dag}).\label{SJT}
\end{eqnarray}
Using Eq. (\ref{SJT}) we obtain the following effective Hamiltonian
\begin{eqnarray}
\hat{H}_{\rm eff}&=&\hbar\omega(\hat{a}_{x}^{\dag}\hat{a}_{x}+\hat{a}_{y}^{\dag}\hat{a}_{y})-\hbar\Omega_{x}\sigma_{x}
-\hbar\Omega_{y}\sigma_{y}-2{\rm i}\frac{\hbar g^{2}}{\omega}\notag\\
&&\times\sigma_{z}(\hat{a}_{x}\hat{a}_{y}^{\dag}-\hat{a}_{x}^{\dag}\hat{a}_{y})-\frac{2\hbar g^{2}}{\omega}-\frac{z_{\rm t}^{2}|\vec{F}_{\perp}|^{2}}{4\hbar\omega}+\hat{H}^{\prime},\label{HeffJTA}
\end{eqnarray}
where $\Omega_{x,y}=g z_{\rm t}F_{x,y}/\hbar\omega$ are the respective Rabi driving frequencies. The next higher-order terms in $\hat{H}^{\prime}$ (\ref{HeffJTA}) are given by
\begin{eqnarray}
\frac{1}{3}[[\hat{H}_{\rm int},\hat{S}],\hat{S}]&=&2{\rm i}\frac{g^{2}z_{\rm t}F_{x}}{\omega^{2}}\sigma_{z}(\hat{a}_{y}^{\dag}-\hat{a}_{y})-2{\rm i}\frac{g^{2}z_{\rm t}F_{y}}{\omega^{2}}\sigma_{z}(\hat{a}_{x}^{\dag}-\hat{a}_{x})\notag\\
&&-\frac{4\hbar g^{3}}{\omega^{2}}\sigma_{y}\{(\hat{a}_{y}^{\dag}+\hat{a}_{y})(1+2\hat{n}_{x})-2\hat{a}_{x}^{\dag 2}\hat{a}_{y}\notag\\
&&-2\hat{a}_{x}^{2}\hat{a}_{y}^{\dag}\}-\frac{4\hbar g^{3}}{\omega^{2}}\sigma_{x}
\{(\hat{a}_{x}^{\dag}+\hat{a}_{x})(1+2\hat{n}_{y})\notag\\
&&-2\hat{a}_{y}^{\dag 2}\hat{a}_{x}-2\hat{a}_{y}^{2}\hat{a}_{x}^{\dag}\}.
\end{eqnarray}

\end{document}